\documentclass[aps,prc,onecolumn,groupedaddress,showpacs,showkeys]{revtex4}
\usepackage[dvips]{graphicx}
\usepackage{bm}
\usepackage{epsfig}
\usepackage{amsfonts}
\usepackage{amssymb,amscd}
\usepackage{subfigure}
\usepackage{latexsym}
\usepackage{amsmath}
\usepackage{slashed}

\newcommand{\be}{\begin{equation}}
\newcommand{\ee}{\end{equation}}
\newcommand{\bea}{\begin{eqnarray}}
\newcommand{\eea}{\end{eqnarray}}
\newcommand{\bem}{\begin{multline}}
\newcommand{\eem}{\end{multline}}
\newcommand{\beg}{\begin{gather}}
\newcommand{\eeg}{\end{gather}}

\newcommand{\ben}{\begin{eqnarray*}}
\newcommand{\een}{\end{eqnarray*}}

\graphicspath{{Figs/}}

\begin{document}

\title{$D$ - meson production at very forward rapidities: estimating the intrinsic charm contribution.}

\author{F. Carvalho$^\dag$, A.V. Giannini$^\ddag$,  V.P.  Gon\c{c}alves$^\S$ and F.S. Navarra$^\ddag$ }
\affiliation{\dag\  Universidade Federal de S\~ao Paulo\\
CEP 01302-907, S\~{a}o Paulo, Brazil}
\affiliation{\ddag\ Instituto de F\'{\i}sica, Universidade de S\~{a}o Paulo\\
C.P. 66318,  05315-970 S\~{a}o Paulo, SP, Brazil}
\affiliation{\S\ Instituto de F\'{\i}sica e Matem\'atica,  Universidade Federal de Pelotas\\
Caixa Postal 354, CEP 96010-900, Pelotas, RS, Brazil\\}

\begin{abstract}
We study $D$ - meson production at forward rapidities taking into account the 
non - linear effects in the QCD dynamics and the intrinsic charm  component  
of the proton wave function. The total cross section, the rapidity distributions  
and the Feynman - $x$ distributions are calculated for $p p$ collisions at different 
center of mass energies. Our results show that, at the LHC, the intrinsic charm 
component  changes  the $D$ rapidity distributions  in a region  which is beyond the
coverage of the LHCb detectors.  At higher energies the IC component dominates 
the y and $x_F$ distributions exactly in the range where the produced $D$ mesons decay 
and contribute the most to the prompt atmospheric neutrino flux measured by the 
ICECUBE Collaboration.  
We compute the $x_F$ - distributions and demonstrate that they are  enhanced at LHC energies by  
approximately one order of magnitude in the $0.2 \le x_F \le 0.8$ range. 
\end{abstract}
\keywords{Intrinsic charm, Particle production, Color Glass Condensate Formalism}
\maketitle
\vspace{1cm}

\section{Introduction}

The  production of charm mesons in $pp/pA/AA$ collisions is an important part of the physics program at the LHC and future colliders \cite{review_hq}. 
One of the reasons to study charm quark production is that it  is expected to be sensitive to the  non - linear effects of  the  QCD dynamics 
\cite{kt_hq,hq_sat,cgn,wata,armesto}, which are predicted to be enhanced at forward rapidites. Another reason is that the understanding of  open charm    
meson production is fundamental to estimate the magnitude of the prompt neutrino 
contribution to the atmospheric neutrino flux \cite{vicmag, enberg,moch,anna,laha}.  
The latter is an important background for the  astrophysical neutrino flux that can be measured by ICECUBE \cite{review_icecube}.  
As demonstrated e.g. in Ref. \cite{anna,laha} and recently discussed in detail in Ref. \cite{vicantoni}, the main contribution to the prompt 
neutrino flux comes from  open charm  meson production at very forward rapidities, 
beyond that reached at the LHC, where new effects may be present.

One of the possible new effects that can contribute to open heavy meson production at  
forward rapidities is the presence of intrinsic heavy quarks in the hadron 
wave function (For recent reviews see, e.g. 
Refs. \cite{nnpdf,hobbs16,review_brod_adv,review_brod_prog}). Heavy quarks in the sea of the proton can be perturbatively generated by gluon splitting.  
Quarks generated in this way are usually denoted  {\sl extrinsic} heavy quarks. 
In contrast, the {\sl intrinsic} heavy quarks have multiple connections to the valence 
quarks of the proton and thus are sensitive to its nonperturbative structure.
Most of the charm content of the proton sea is extrinsic and comes from the 
DGLAP \cite{dglap} evolution of the initial gluon distribution. This process is well understood in perturbative QCD.  The existence of the intrinsic charm (IC) component was 
first proposed long ago in Ref. \cite{bhps} (see also Ref. \cite{hal}) and since 
then other models for IC  have been  discussed.

In the original model \cite{bhps,vb}, the creation of the $c \overline{c}$ pair was 
studied in detail. It was assumed 
that the nucleon light cone wave function has  higher Fock states,  one of them being
$|q q q c \overline{c}>$. The probability of finding the nucleon in this
configuration is given by the inverse of the squared invariant mass of the
system. Because of the heavy charm mass, the probability distribution as a function 
of the
quark fractional momentum, $P(x)$, is very hard, as compared to the one obtained
through the DGLAP evolution. In the literature this model is known as BHPS.
A more dynamical approach is given by the meson cloud model (MC). In this model, the nucleon fluctuates into an intermediate state 
composed by a charmed baryon plus a charmed meson \cite{pnndb,wally99}. The charm is always confined in one hadron and carries the largest part of its momentum. 
In the 
hadronic description we can use effective lagrangians to compute the charm splitting functions, which turn out to favor harder charm quarks than the DGLAP ones.
The main difference between the BHPS and MC  models is that the latter predicts
that  the charm and  anticharm distributions are different \cite{cdnn}, since they carry 
information about the hadronic bound states in which  the quarks are. 

%The  {\sl intrinsic} charm  component has been studied in 
%detail and  included in  some versions of the CTEQ parameterization \cite{cteq}. 
%The hypothesis of intrinsic heavy quarks (IHQ) is a natural consequence of the quantum fluctuations 
%inherent to Quantum Chromodynamics (QCD) and amounts to assuming the existence of a $Q\bar{Q}$ ($Q = c,b,t$) 
%as a nonperturbative component in the hadron wave function, which can be realized through an 
%interaction.  One  of the most striking properties of an IHQ state, such has $|uudQ\bar{Q}\rangle$, 
%is that the heavy constituents tend to carry the largest fraction of the momentum of the hadron. 
%Although the existence of intrinsic charm in the proton has some  experimental 
%and theoretical support, as discussed in  Refs. \cite{bv15,pumplin}, 
%more definite conclusions are  not yet possible. 
%The idea of intrinsic charm was left aside for some years.  

\begin{figure}[t]
 \begin{center}
 \includegraphics[width=0.6\textwidth]{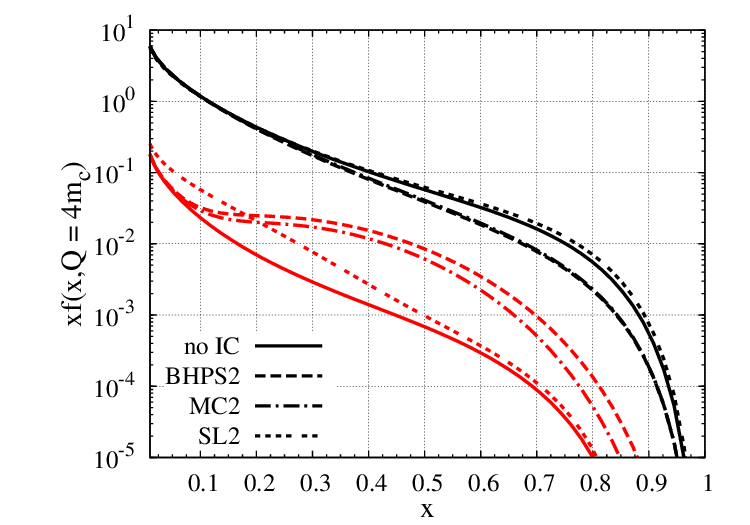}
 \caption{Predictions of the different intrinsic charm models for the $x$ - dependence of the  charm (lower red curves) and gluon (upper black curves) 
distributions as obtained by the CTEQ Collaboration \cite{cteq}.}
  \label{Fig_pdf}
\end{center}
\end{figure}

The  intrinsic charm (IC) component of the  proton wave function was considered in 
the global analysis  performed in 2006 by the 
CTEQ collaboration \cite{cteq}. In this update the CTEQ group 
determined the  normalization of the IC distribution. 
In fact, they find several IC distributions 
which were compatible with the world data. Apart from
the already mentioned BHPS and meson cloud models, the CTEQ group has tested 
another model of intrinsic charm, called sea-like IC. It consists basically in 
assuming that at a very low resolution (before the DGLAP evolution) there is 
already some charm in the nucleon, which has a
typical sea quark momentum distribution ($\simeq 1/ \sqrt{x}$) with normalization 
to be fixed by fitting data. The resulting charm distributions are presented in Fig. 
\ref{Fig_pdf} (lower red curves), where the no IC curve represents the standard CTEQ prediction, obtained disregarding the presence 
of  intrinsic charm in the initial condition of the evolution. The BHPS and MC models predict a large enhancement of the distribution 
at large  $x$  ($> 0.1$). In contrast, the sea - like (SL) model predicts a smaller enhancement at large  $x$, but a larger one at 
smaller  $x$ ($< 0.2$). We follow Ref. \cite{cteq} and use the labels BHPS2, MC2 and SL2 for 
the versions of these models  which have the maximum amount of intrinsic charm.
In Fig. \ref{Fig_pdf} we also present the corresponding gluon distributions (upper black curves). Due to the 
momentum sum rule, the gluon distribution is also  modified by the inclusion of  intrinsic charm. In particular, the BHPS and MC models imply a 
suppression in the gluon distribution at large  $x$ (For a more detailed discussion see e.g. Ref. \cite{iranianos}).

%As we will show below, 
%this modification in $xg$ has important implications in the $D$ - meson production at forward rapidities.

The large enhancement  at large - $x$ in the charm distribution, associated to  intrinsic charm, has motivated a large number of phenomenological studies 
to confirm the presence (or absence) of this component in the hadron wave function. One of the most  direct consequences is that the intrinsic charm component 
gives rise to heavy mesons with large fractional momenta relative to the beam particles,  affecting  the  Feynman - $x$ ($x_F$) and rapidity distribution of 
charmed particles. This aspect was explored e.g. in  Refs. \cite{ingelman,kniehl,npanos}). Moreover, the presence of  intrinsic charm changes the Higgs   
\cite{brod_higgs} and photon production \cite{russos} at high $x_F$. Over the last two years,  new parametrizations of the IC distribution were released 
\cite{cteq14,wally16}, motivating an intense debate about  the amount of IC  in the proton wave function \cite{wally16,brod_letter}. At the same time new 
implications of IC were discussed \cite{bailas,biw16,icbot}. In particular, in Ref. \cite{icbot} the authors presented a method to generate  matched  
intrinsic charm / intrinsic bottom distributions for any PDF set without the need for a complete global re-analysis. This allows one to easily carry out a 
consistent analysis including intrinsic quark effects. Additionally, the proposal of constructing a  
high energy and high luminosity fixed-target experiment 
using the LHC beams (AFTER@LHC) \cite{after}  motivated new theoretical studies about the possibility of measuring the intrinsic charm component of the nucleon 
\cite{review_brod_adv}. Finally, the effect of IC on the atmospheric neutrino flux measured by ICECUBE was 
addressed in Refs. \cite{halzen,laha,rojo_ice} considering different phenomenological models for the treatment of the intrinsic component.

In this paper we revisit  $D$ - meson production at forward rapidities in hadronic collisions, which probes particle production at  large $x_F$. In this case,  
the kinematics is very asymmetric, with the hadrons in the final state emerging from 
collisions of projectile partons with large light cone momentum 
fractions  with target partons carrying  a very small 
momentum fraction. As a consequence,  we have the scattering of a 
dilute projectile on  a dense target, where the small-$x$ effects coming from the non-linear aspects of QCD and from the physics of the Color Glass Condensate 
(CGC)  \cite{cgc} are expected to appear and the usual factorization formalism is expected to breakdown \cite{hq_sat}. 
The satisfactory description of the experimental data in this 
kinematical region with the CGC approach \cite{dhj,buw,ptmedio.nois} indicates 
that the CGC is the appropriate framework to study particle production in the large rapidity region (for an alternative approach, see \cite{rojo_prl}).
Along this line, in \cite{npanos} the formalism 
proposed in \cite{dhj} for light meson production  was generalized  to  $D$-meson production in $pp$ and $p A$
collisions at forward rapidities, including  intrinsic charm quarks in the 
projectile wave function. Recently, the basic equation proposed in Ref. \cite{npanos} was reobtained  in Ref. \cite{armesto}. In Ref. \cite{npanos} we 
have presented predictions for  the $p_T$ distributions of $D$ mesons at large rapidities considering $pp$ and $pA$ collisions at RHIC and LHC energies. 
Our results indicated that the presence of  intrinsic charm strongly modifies the $p_T$ - spectra. However, a shortcoming of Ref. \cite{npanos} is that the 
gluonic  contribution to $D$ - meson production, associated to the $g + g \rightarrow c + \bar{c}$ channel, was not included in a systematic way. Basically, 
this contribution was considered as a background and was estimated considering the standard PYTHIA predictions. One of the main goals of this paper is to 
consistently include  this contribution, taking into account the non - linear effects  in the QCD dynamics at small  $x$, as well as the modifications at 
large  $x$ in the gluon distribution predicted by the different IC models. We will present predictions for the total cross section and rapidity distribution 
considering $pp$ - collisions at LHC energies and compare them with the recent experimental data. In particular, we will estimate the rapidity region where 
the IC contribution is larger.  Another goal is to estimate the impact of the IC component on the  $x_F$ - distributions, which are the main input in the 
calculations of the prompt neutrino flux. We will present our predictions for this distribution considering LHC and Ultra - High Cosmic Rays (UHECR) energies  
and demonstrate that the $x_F$ behavior is strongly modified in the $x_F \ge 0.2$ range. 

This paper is organized as follows. In next Section we will present a brief review of the main ingredients used in our calculation of  $D$ - meson production 
at forward rapidities and large  $x_F$. In particular, we review the approach proposed in Ref. \cite{npanos} for the intrinsic component and the dipole picture  
of  heavy quark production in gluonic interactions developed in Refs. \cite{nnz,boris} and applied at the LHC in Refs. \cite{cgn,vicboris}.  Both contributions will be 
expressed in terms of the  dipole - nucleon scattering amplitude, which we will assume to be given by the model proposed in Ref. \cite{buw} and recently updated 
to describe the recent LHC data on forward particle production in $pp$ collisions in Ref. \cite{ptmedio.nois}.   
In Section \ref{res} we present our results for the total cross section and rapidity distribution and compare with recent experimental data. The impact of the 
IC contribution is estimated and the optimal kinematical range to probe its presence is determined. Moreover, we estimate the $x_F$ - distribution considering 
$pp$ collisions at LHC and  UHECR energies. Finally, in Section \ref{conc} we summarize our main conclusions.

%In this approach the target is treated as a dense 
%system  while the projectile proton is taken to be a dilute system in the spirit 
%of standard QCD. The charm quarks were assumed to be already in the projectile with  
%their densities  given by the CTEQ parametrization \cite{cteq}. 
%The interaction of these charm quarks were described in terms of the dipole 
%scattering amplitude  proposed in \cite{buw}. After the interaction with the target 
%the charm quarks fragment into $D$ mesons according to the fragmentation
%functions given in Ref. \cite{bkk}. 

\section{$D$ - meson production at forward rapidities}

In what follows we will address prompt $D$ - meson production, disregarding the contribution from the decay of heavier mesons. In this case, the $D$ - meson 
production is determined by the cross section of  heavy quark production, which is  usually described in the collinear or $k_T$-factorization 
frameworks of QCD. Using  collinear factorization, charm production  
is described  in terms  of the  basic subprocesses of gluon fusion 
($ g + g \rightarrow c + \overline{c}$) and light quark-antiquark fusion   
($ q +  \overline{q} \rightarrow c + \overline{c}$), with the latter being negligible at high energies. The elementary cross section computed 
to leading order (LO) or next-to-leading order (NLO) is folded with the corresponding parton 
distributions and fragmentation functions. This is the basic procedure in most of the 
calculations performed, for instance,  with the standard codes PYTHIA and MCFM. However, at small $x$  collinear factorization should be generalized to 
resum powers of $\alpha_s \ln (s/q_T^2)$, where $q_T$ is the transverse momentum of the final state and 
$\sqrt{s}$ is the center of mass energy. This resummation is done in the $k_T$-factorization framework, where the cross section is expressed in terms of 
unintegrated gluon distributions which are determined by the QCD  dynamics at small-$x$ (For recent results see e.g. Ref. \cite{antoni}). The presence of 
non - linear effects in the QCD dynamics is expected to have impact on  heavy quark production at high energies \cite{kt_hq,hq_sat,cgn,wata}, leading to the 
breakdown of the $k_T$-factorization \cite{hq_sat, marquet}.

\begin{figure}[t]
 \begin{center}
 \includegraphics[width=0.3\textwidth]{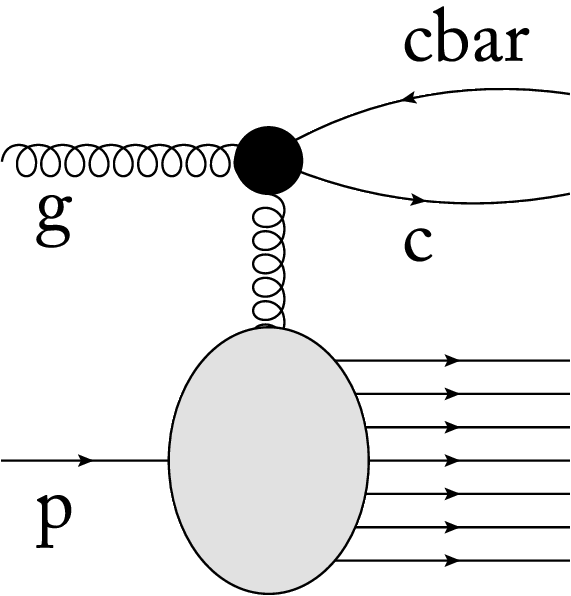} \hspace{2cm}
 \includegraphics[width=0.3\textwidth]{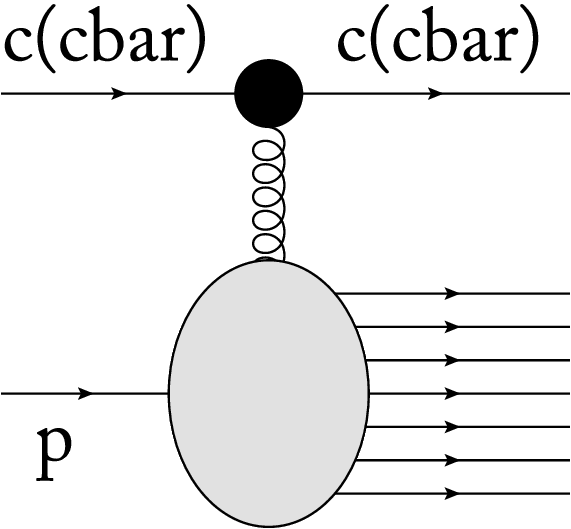}
  \end{center}
%  \vskip -0.80cm
 \caption{Contributions to $D$ - meson production at forward rapidities. Left panel:  Contribution from gluon - gluon interactions. Right panel: 
Contribution from charm in the initial state.}
  \label{Fig:diagrama}
\end{figure}

One way to study heavy quark production in gluon - gluon interactions is the color dipole formalism developed in Refs. \cite{nnz,boris}, which allows us to 
take into account the non - linear effects in the QCD dynamics as well as higher order corrections \cite{rauf}. The basic idea  of  this approach is that, before 
interacting with the hadron target, a gluon is emitted by the projectile and fluctuates into  a color octet pair $Q\bar{Q}$. The dipole picture of HQ production is 
represented in  Fig. \ref{Fig:diagrama} (left panel). Taking into account that the heavy quarks in the dipole as well the incident gluon (before fluctuating into 
the pair) can interact with the target, the rapidity distribution for a $h_1 h_2$ collision can be expressed as follows \cite{boris}:
\begin{equation}
\frac{d\sigma(h_1 h_2 \rightarrow \{ Q\bar{Q} \}X)}{dy} =  x_1G^{h_1}(x_1,\mu_F^2) \,
\sigma (Gh_2 \rightarrow \{Q\bar{Q} \}X)\,\,,
\label{dsdy_gluons}
\end{equation}
where $x_1G_{p}(x_1,\mu _F)$ is the projectile gluon distribution, the cross section  $\sigma (Gh_2 \rightarrow \{Q\bar{Q}\} X)$ describes  
heavy quark production in a gluon - nucleon interaction, $y$  is the rapidity of the pair and $\mu_F$ is the 
factorization  scale.  Moreover,  the cross section of the process  $G + h_2 \rightarrow Q \bar{Q} X$ is given by:
\begin{equation}
\sigma(G h_2 \rightarrow\{Q\bar{Q}\}X) = \int _0^1 d \alpha \int d^2\rho \,\,
\vert \Psi _{G\rightarrow Q\bar{Q}} (\alpha,\rho)\vert ^2
\,\, \sigma^{h_2} _{q\bar{q}G}(\alpha , \rho)
\label{sec1}
\end{equation}
where  $\alpha$ ($\bar{\alpha} \equiv 1 - \alpha$) is the longitudinal momentum fraction carried by the quark (antiquark), $\vec{\rho}$ is the transverse 
separation of the pair,   $\Psi _{G\rightarrow Q\bar{Q}}$ is the 
light-cone (LC) wave function of the 
transition $G \rightarrow  Q \bar{Q} $ (which is calculable perturbatively and is proportional to $\alpha_s$)  and $  \sigma^{h_2}_{Q\bar{Q}G}$  is the 
scattering cross section of a color neutral quark-antiquark-gluon system on the 
hadron target $h_2$ \cite{nnz,boris,rauf}. The three - body cross section is given in terms of the dipole - nucleon cross section $\sigma _{q\bar{q}}$ as follows:
\begin{equation}
\sigma^{h_2}_{q\bar{q}G}(\alpha , \rho) = \frac{9}{8}[\sigma _{q\bar{q}}(\alpha \rho) + \sigma _{q\bar{q}}(\bar{\alpha} \rho)]
- \frac{1}{8}\sigma _{q\bar{q}}(\rho)\,\,.
\label{sec2}
\end{equation}
The dipole - nucleon cross section can be expressed in terms of the forward scattering amplitude ${\cal{N}} (x,\rho)$, which is determined by the QCD dynamics 
and constrained by the HERA data, as follows:
\begin{equation}
\sigma_{q \bar q}(x,\rho) = \sigma_{0} {\cal N}(x,\rho)
\label{sigzero}
\end{equation} 
where $\sigma_0$ is a free parameter usually determined by a fit of the HERA data. 
In the dipole picture  the heavy quark production cross section is associated to gluon - gluon interactions and it is determined by the 
projectile gluon distribution and by the model assumed for the dipole - nucleon scattering amplitude. Moreover, it depends on the values of the charm mass and 
of the running coupling constant $\alpha_s$. Finally, in order to estimate the corresponding $D$ - meson cross section we need to convolute the heavy quark cross 
section with  the fragmentation function, for which  a model must be chosen.

As discussed in the Introduction,  the cross sections at forward rapidities are dominated by collisions of projectile partons with large light cone momentum 
fractions  with target partons carrying  a very small 
momentum fraction. From light hadron production, 
we know \cite{barstop} that in this kinematical range the cross section is
dominated by the interaction of valence quarks of the projectile with gluons of the target.  In other words, the cross 
section  depends on the partonic structure of the projectile at large-$x$. 
If  the intrinsic charm is present in the proton wave function and strongly modifies  the behavior of the corresponding parton distribution at large $x$, 
it is natural to expect that IC may change the $D$-meson production cross section. Additionally, as we are probing very small values of $x$ in the target,  
non - linear effects in QCD dynamics should be taken into account. These were the basic motivations of the study performed in Ref. \cite{npanos}, where we 
generalized the DGLAP $\otimes$ CGC factorization scheme proposed in Ref. \cite{dhj} to estimate 
the intrinsic charm contribution (For a recent derivation see Ref. \cite{armesto}).  In this approach the projectile (dilute system) evolves according to the 
linear DGLAP dynamics and the target (dense system) is treated using the CGC formalism. As a consequence the differential cross section of  $D$ - meson 
production associated to  charm in the initial state is given by \cite{npanos}
\begin{eqnarray}
{d\sigma \over dy d^2p_T} = 
{1 \over (2\pi)^2} \int_{x_F}^{1} \frac{dz}{z^2}
f_{c/p}(x_1,q_T^{2})\,\, \sigma_0 \, \widetilde{\cal N} \left(x_2, {p_T \over z}\right)\,
D_{D/c}\left(z,\mu^{2}_{FF}\right)
\label{dNdy_quarks}\,.
\end{eqnarray}
with the  
variables  $x_{F}$ e $x_{1,2}$ being  defined by $x_{F} = \sqrt{p_{T}^{2} + m^{2}}\, e^{y}/\sqrt{s}$ and
$x_{1,2} = q_{T} e^{\pm y}/\sqrt{s}$, where $q_{T} = p_{T}/z$.
Therefore,  particle production at forward rapidities and small values of transverse momentum is characterized by the interaction between partons with 
large $x_1$ in the projectile and small values of $x_2$ in the target. As a consequence, the hadron in the final state is expected to be produced at 
large values of $x_F$. Moreover, $f_{c/p}$ represents the projectile charm distribution, $D_{D/c}$ is the charm 
fragmentation function in a $D$ - meson and $\widetilde{\cal N} (x, {k_T})$ is the Fourier transform of the scattering amplitude ${\cal N}(x,\rho)$. 
The basic diagram associated to this process is presented in Fig. \ref{Fig:diagrama} (right panel).  In Ref. \cite{npanos} we 
estimated the $p_T$ - spectra of the $D$ - mesons produced at different rapidities in $pp$ and $pA$ collisions at RHIC and LHC energies and demonstrated 
that the  IC component in the proton wave function implies a strong enhancement of the differential cross sections in comparison with the predictions derived 
disregarding this component. In the next Section we will extend the analysis performed in Ref. \cite{npanos}, calculating the corresponding rapidity and 
$x_F$ - distributions and including the contribution of gluon-gluon interactions to $D$-meson production. In order to be 
theoretically consistent, we will estimate Eqs.  (\ref{dsdy_gluons}) and (\ref{dNdy_quarks})  using a common pdf set for the gluon and  charm distributions 
as well the same models for the scattering amplitude and fragmentation function.

\section{Results and discussion}
\label{res}

In what follows we will present our predictions for  $D$-meson production at forward rapidities and large - $x_F$ considering the two contributions discussed 
in the previous Section. This kinematical region is characterized by large values of $x_1$ and small $x_2$, which implies an asymmetric projectile -- target 
configuration, usually denoted dilute -- dense one. It is exactly for this configuration that the dipole approaches for  heavy quark production considered in 
our analysis have been derived \cite{boris,dhj,npanos}.  In order to calculate  the cross sections we need to specify the parametrization used for the parton 
distributions, the model for the scattering amplitude and the $c \to D$ fragmentation function. Moreover, the results  depend on the charm mass, on the 
factorization scale and on the running 
coupling constant.

Let us start discussing the model assumed for the scattering amplitude 
${\cal N}(x,\rho)$ and, consequently, for $\widetilde{\cal N} (x, {k_T})$. This quantity involves the QCD dynamics at high energies and contains all the information 
about the initial state of the hadronic wavefunction and therefore about the non-linearities and quantum effects which are characteristic of a system such as the 
CGC (For reviews, see e.g. \cite{cgc}). Formally its evolution is usually described in the mean field approximation of the CGC formalism by the BK equation \cite{bk}. 
Its analytical solution is known only in some special cases.  Advances have been made in solving the BK equation numerically 
\cite{la11}. Since the BK equation still lacks a formal solution in all phase space, several groups have constructed phenomenological models for the dipole 
scattering amplitude. These models have been used to fit the RHIC and HERA data \cite{dhj,buw,dips}. In general, it is  assumed that
${\cal{N}}$ can  be modelled  through a simple Glauber-like formula,
\begin{eqnarray}
{\cal{N}}(x,\rho) = 1 - \exp\left[ -\frac{1}{4} (\rho^2 Q_s^2(x))^{\gamma (x,\rho^2)} \right] \,\,,
\label{ngeral}
\end{eqnarray}
where $Q_s(x)$ is the saturation scale, $\gamma$ is the anomalous dimension of the target 
gluon distribution and $\rho$ is  the dipole size.  The speed with which we move from the non-linear 
regime to the extended geometric scaling regime and then from the latter to the linear regime is what differs one phenomenological model from another. This transition 
speed is dictated by the behavior of the anomalous dimension $\gamma (x,\rho^2)$. In this paper we consider the BUW \cite{buw} dipole model, which assumes that the 
anomalous dimension can be expressed by
\begin{equation}
\label{dip_adimension}
\gamma_{BUW} = \gamma_{s} + (1-\gamma_s)\frac{(\omega^a-1)}{(\omega^a-1)+b}
%\label{anodim}
\end{equation}
where $\omega = q_T/Q_s$ and $a$, $b$ and $\gamma_s$ are free parameters to be fixed 
by fitting  experimental data. In the BUW Ansatz, the anomalous dimension $\gamma$ leads to 
geometric scaling and hence depends only on the ratio $\omega = q_T/Q_s$ but not separately 
on $q_T$ and on $Q_s(x)$. 
Recently, in Ref. \cite{ptmedio.nois},  the original parameters of the 
BUW model  were updated in order to  make this model compatible with all existing data. In particular, the recent LHC data on light hadron production at forward 
rapidity are satisfactorily reproduced by the updated model. In what follows we will use the BUW model with  the parameters obtained in Ref. \cite{ptmedio.nois}.

In order to quantify the impact of the intrinsic charm considering the largest possible number of models of this component, we will use in our calculations the 
leading order  CTEQ 6.5 parametrization for the  parton distributions \cite{cteq}. This particular parametrization has  two different PDF sets for each of the 
models discussed in the Introduction (BHPS, MC and SL), considering different amounts of intrinsic component. It is important to emphasize that this amount is 
still subject of intense debate. The recent IC global analysis presented in 
\cite{wally16} comes to the conclusion that such a big amount of IC in the proton is excluded by the current experimental data. Depending on the analysis,  the 
obtained upper limit on the IC normalization is around 2.5 \% or  0.5 \%.  As our goal is highlighting the possible effect of IC,  we will only present the 
predictions obtained with the CTEQ 6.5 parametrization with the maximum amount of IC ($\simeq 3.5$ \%) for a given model (denoted BHPS2, MC2 and SL2 hereafter).
   
It is important to emphasize that the CTEQ-TEA group  has also performed a global analysis of the recent experimental data including an intrinsic 
charm component, which is available in the CT14 parametrization \cite{cteq14}. However, this analysis has been performed at next-to-next-to-leading order and the 
meson cloud model was not considered. As the basic equations used in our study of  $D$ - meson production have been derived at leading order, we believe that it 
is more consistent to use in our calculations PDFs obtained at the same order. In order to analyse the influence of a more recent leading order parametrization, 
we  will also present in some of our results, the predictions obtained using the CT14 LO parametrization, which disregards the intrinsic component.
\begin{figure}[t]
 \begin{center}
 \subfigure[ ]{
\includegraphics[width=0.475\textwidth]{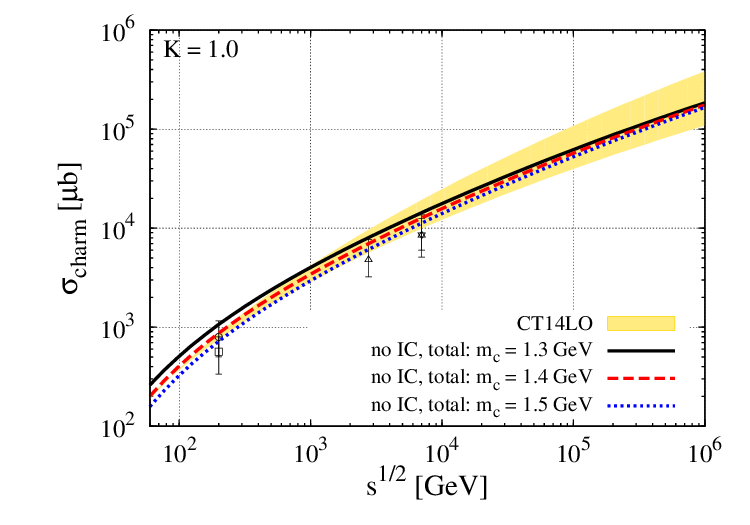}}
\subfigure[ ]{
 \includegraphics[width=0.475\textwidth]{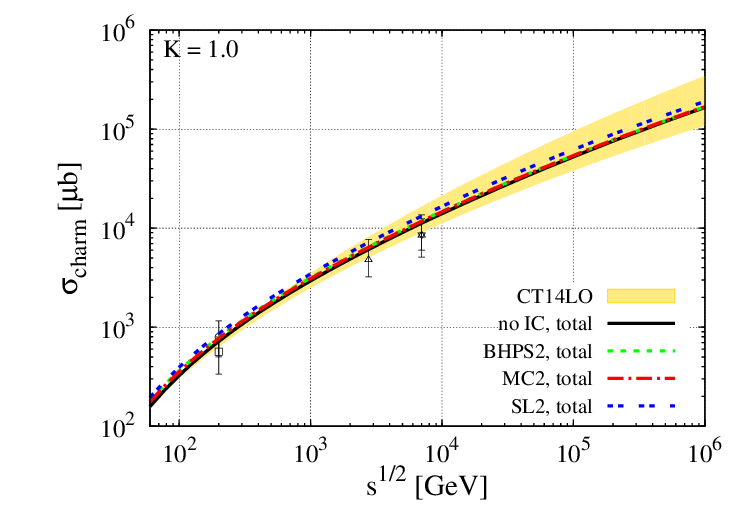}}
  \end{center}
  \vskip -0.80cm
 \caption{Energy dependence of the total charm production cross section considering (a) different values of the charm mass, and (b) different models of the 
intrinsic charm component.  Data from Refs. \cite{hera7,star12,alice12,lhcb11,lhcb13,lhcb16}. }
  \label{Fig_sectot}
\end{figure}
%As we are interested in the total cross section and the rapidity and $x_F$ distributions, which are not strongly affected by the modelling of the fragmentation 
%function,  we will assume  a naive model for the fragmentation function of a charm into a $D$ - meson. Basically, we will assume that 
%$D_{h/c} = f_h \, \delta(1-z)$, where $f_h$ is the fraction of fragmentation of $c \rightarrow h$. For the $D^0$ we take $f_{D^0} =0.565 $, as derived in 
%Ref. \cite{enberg}.
Following Ref. \cite{anna} we will use the  fragmentation function from Ref.~\cite{kk} given by: 
\begin{equation}
\label{eq:frag}
D_c^h(z) = \frac{Nz(1-z)^2}{[(1-z)^2+\epsilon z]^2}\ .
\end{equation}
where  $N=0.577$ and $\varepsilon = 0.101$ and the fragmentation fraction into $D^0$ is $0.606$. An alternative is to consider fragmentation functions with 
DGLAP evolution as  e.g. those obtained in Ref. \cite{kramer}. However, as demonstrated in Ref. \cite{vicboris}, the main implication of DGLAP evolution is  
the modification of the $p_T$ - spectra at large tranverse momenta. In our analysis we are interested in the rapidity distribution, which is dominated by 
small values of $p_T$. Therefore, the DGLAP evolution effects in the fragmentation function are expected to have a negligible impact on  $d\sigma/dy$. We 
also will assume that the factorization scale 
$\mu_F$ in Eq. (\ref{dsdy_gluons}) is equal to $\mu_F^2 = 4 \cdot m_c^2$ and that 
$\alpha_s = \alpha_s(\mu_F^2)$. As a consequence, our predictions depend only on the choice of the  charm mass. Finally, it is important to emphasize that our 
predictions for the gluon and charm contributions, given by the Eqs. 
(\ref{dsdy_gluons}) and (\ref{dNdy_quarks}), could be modified by higher order corrections, which are in several cases mimicked by a $K$ - factor multiplying 
the expressions, fitted to describe the data. In our analysis, we will assume that $K = 1$. 
However, the estimate of higher corrections for the gluonic and charm contributions is a subject that deserves a more detailed study in the future.

In Fig. \ref{Fig_sectot} (a) we show our results for the total charm production cross section,  considering different values of the charm mass, 
summing the gluonic and charm contributions and assuming that the gluon and charm PDFs are given by the standard CTEQ 6.5 parametrization without an 
intrinsic component. We compare our predictions with the experimental data. We can observe that the data at high energies are reasonably well described.  
In what follows we will assume that $m_c = 1.5$ GeV. For comparison we also present the CT14 LO predictions for $m_c = 1.5$ GeV and different values for the 
factorization scale $\mu_F$, in order to estimate the dependence of our predictions on the choice of this scale. Considering the range 
$m_c^2 \le \mu_F^2 \le  16 \, m_c^2$, we observe  that the CT14 LO parametrization leads to the band presented in Fig. \ref{Fig_sectot}, which demonstrates 
that the high energy behavior of the cross section is strongly dependent on the choice of  $\mu_F$.
In Fig. \ref{Fig_sectot} (b) we investigate the impact of the different IC models on the total charm cross section. We can see that the different predictions 
are almost identical, which is expected, since the total cross section is dominated by the contribution associated to  charm production at central rapidities, 
where both $x$ values of the partons involved in the collision are small. As a consequence, the modifications associated to the presence of  intrinsic charm are 
negligible for this observable.

\begin{figure}[t]
 \begin{center}
 \subfigure[ ]{ \includegraphics[width=0.475\textwidth]
%{LHC_pp_dsigdy_7000_NNLO.eps}}
%{dsdy_7tev_NEW.eps}}
{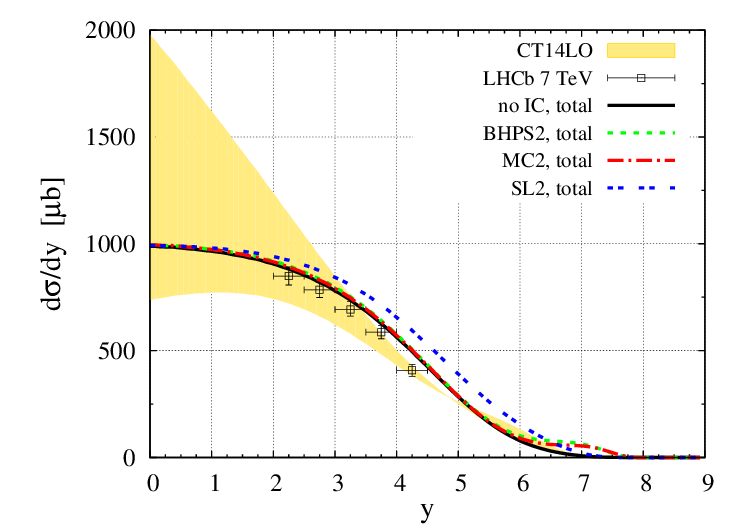}}
  \subfigure[ ]{\includegraphics[width=0.475\textwidth]
%{LHC_pp_dsigdy_7000_BHPS2_.eps}}
%{dsdy_13tev_NEW.eps}}
{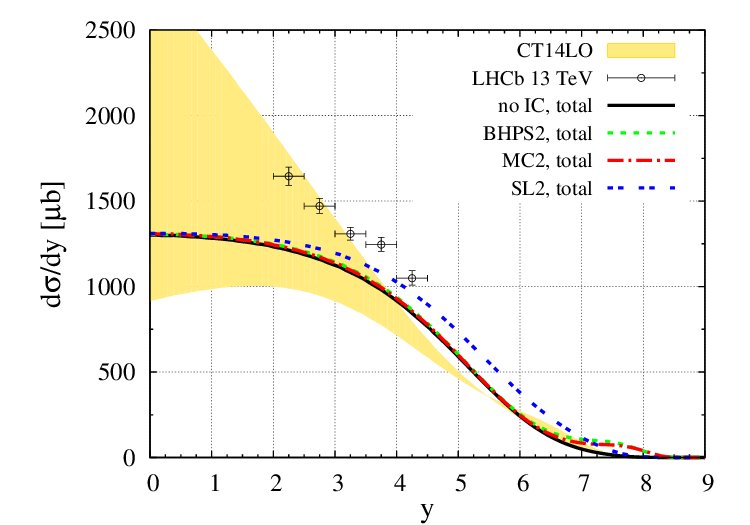}}
  \end{center}
 \vskip -0.80cm
 \caption{Rapidity distribution of  $D^{0}+\bar{D^{0}}$ mesons produced in $pp$ collisions at (a)  $\sqrt{s}=7$ TeV and (b) $\sqrt{s}=13$ TeV. 
Data  from Refs.~\cite{lhcb13,lhcb16}.}
 \label{Fig:dsigdy}
\end{figure}

%\begin{figure}[t]
%\begin{center}
%\includegraphics[width=0.6\textwidth]{R_IC_BHPS2_dsigdy_NEW_scales.eps}
%\caption{Rapidity dependence of the ratio between the IC (BHPS2) predictions and the standard CTEQ 6.5 parametrization without intrinsic charm.}
%\label{Fig:ratio_rap}
%\end{center}
%\end{figure}

\begin{figure}[t]
 \begin{center}
 \subfigure[ ]{ \includegraphics[width=0.475\textwidth]
%{LHC_pp_dsigdy_7000_NNLO.eps}}
%{R_IC_BHPS2_dsigdy_NEW_scales.eps}}
{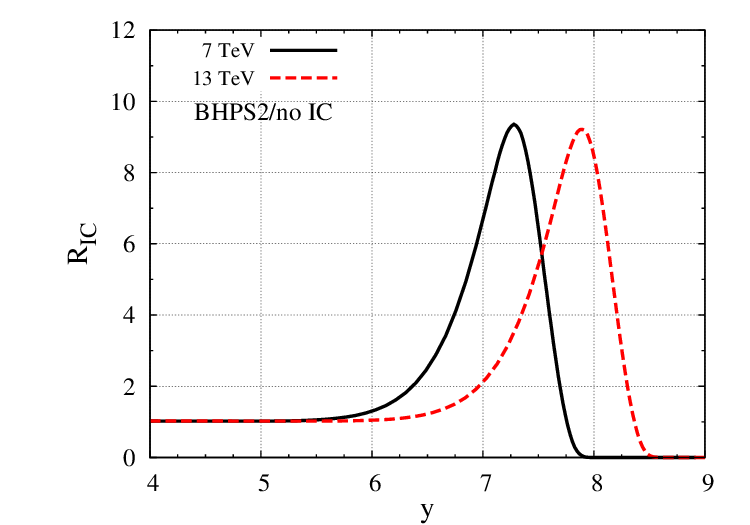}}
  \subfigure[ ]{\includegraphics[width=0.475\textwidth]
%{LHC_pp_dsigdy_7000_BHPS2_.eps}}
%{R_IC_MC2_dsigdy_NEW_scales.eps}}
{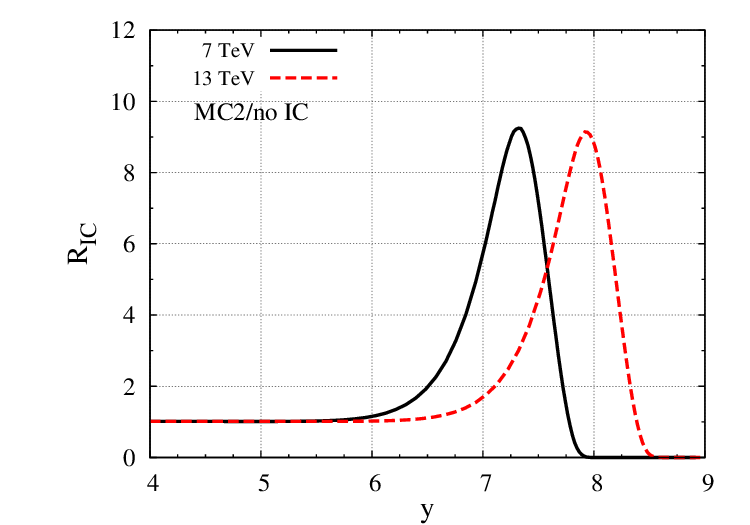}}
  \end{center}
 \vskip -0.80cm
\caption{Rapidity dependence of the ratio between the IC  predictions and the standard CTEQ 6.5 parametrization without intrinsic charm: a) BHPS2. 
b) MC2.}
\label{Fig:ratio_rap}
\end{figure}

Our predictions for the rapidity distribution of $D^{0}+\bar{D^{0}}$ mesons, produced in $pp$ collisions at $\sqrt{s} = 7$ and 13 TeV, are presented in 
Figs. \ref{Fig:dsigdy} (a) and (b), respectively. Considering the CT14 LO predictions, we can see that the predictions at central rapidities  are strongly 
sensitive to the choice of $\mu_F$, with the uncertainty decreasing at larger rapidities,  which is the kinematical range where the IC component contributes. 
The standard CTEQ 6.5 and the  BHPS2 and MC2 models  predict a similar 
behavior at small rapidities, differing only at very forward rapidities. On the other hand, the SL2 model predicts an enhancement in the rapidity distribution 
in the region of intermediate values of $y$, which is directly associated to the enhancement of the charm distribution for $x \le 0.2$ (See Fig. \ref{Fig_pdf}). 
In order to quantify the influence of the intrinsic component and determine the kinematical region affected by its presence, in Fig.    
\ref{Fig:ratio_rap} we present the rapidity dependence of the ratio between the IC predictions and the standard CTEQ 6.5 one without an intrinsic component 
(denoted no IC). Our results demonstrate that the intrinsic component modifies the rapidity distribution at very forward rapidities, beyond those reached by 
the LHCb Collaboration.  We observe that the distribution can be enhanced by a factor $\approx 8$, with the position of the maximum    shifting to larger 
rapidities with the growth of the center of mass energy. This behavior can be easily understood if we remember that the $x$ - values probed in the projectile 
are approximately $x \approx (m_T/\sqrt{s})\exp(+y)$, where $m_T = \sqrt{4 m_c^2 + p_T^2}$. Consequently, when the energy increases, we need to go to larger 
rapidities in order to reach the same value of $x$.

\begin{figure}[t]
 \begin{center}
 \subfigure[ ]{ \includegraphics[width=0.475\textwidth]
%{LHC_pp_dsigdy_7000_NNLO.eps}}
%{termos_noic_13tev_NEW.eps}}
{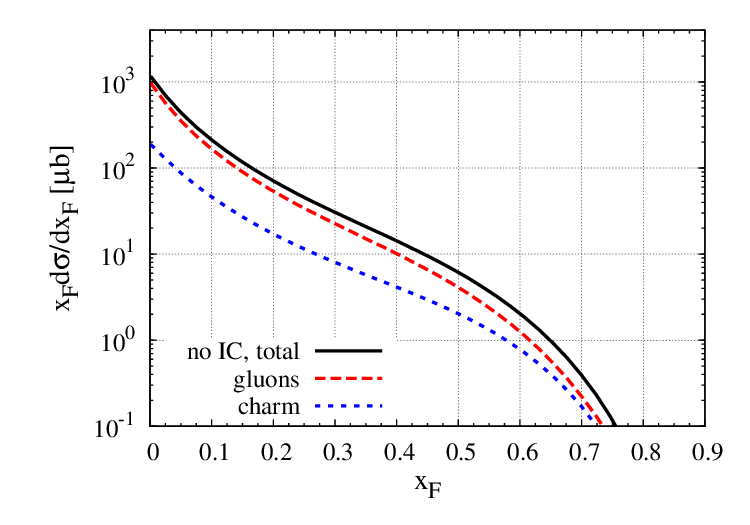}}
  \subfigure[ ]{\includegraphics[width=0.475\textwidth]
%{LHC_pp_dsigdy_7000_BHPS2_.eps}}
%{termos_bhps_13tev_NEW.eps}}
{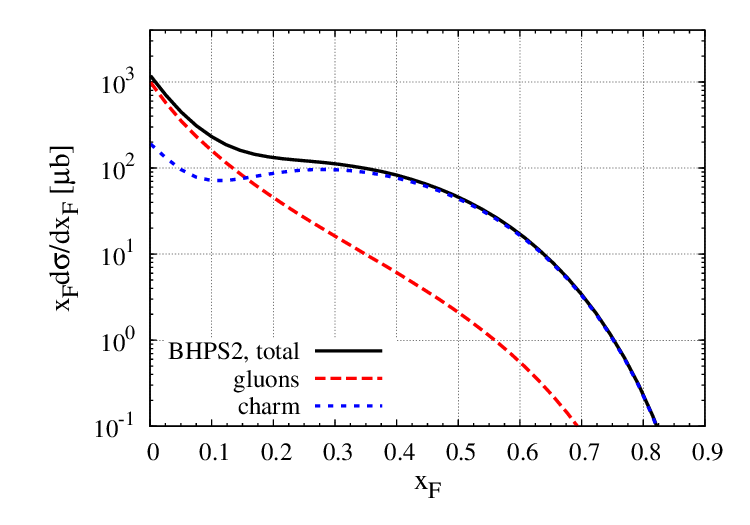}}
  \end{center}
 \vskip -0.80cm
 \caption{Feynman - $x$ distributions  of the produced   $D^{0}+\bar{D^{0}}$ mesons in $pp$ collisions at $\sqrt{s} = 13$ TeV considering: 
(a) the standard CTEQ 6.5 parametrization and 
(b) the BHPS model. The gluonic and charm contributions are presented separately.}
 \label{Fig:termos}
\end{figure}

\begin{figure}[htb]
 \begin{center}
  \subfigure[ ]{\includegraphics[width=0.475\textwidth]
%{LHC_pp_xfdNdxf_7a_NEW_scales.eps}}
{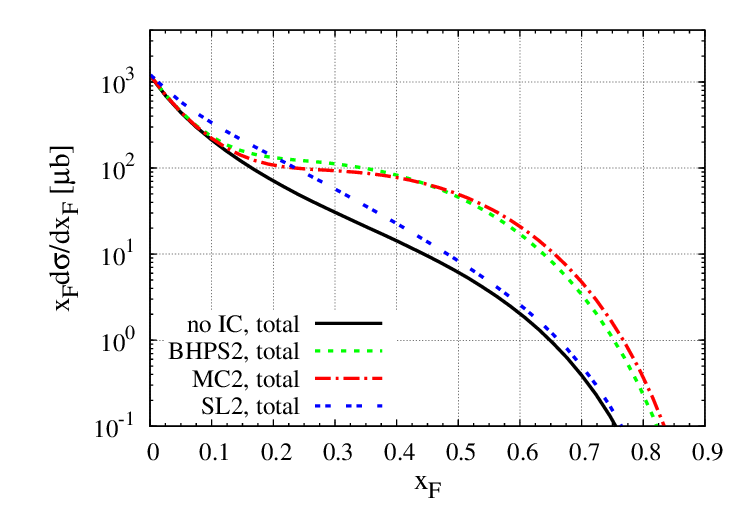}}
\subfigure[ ]{\includegraphics[width=0.475\textwidth]
%{dsdxf_total_200tev_NEW.eps}}
{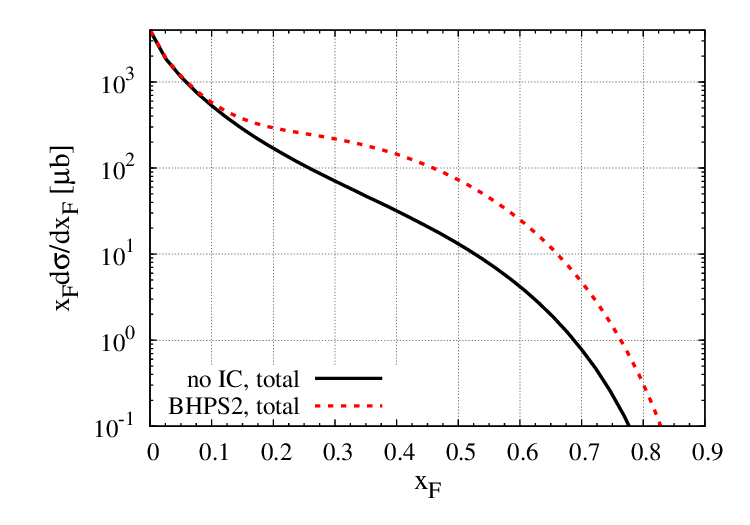}}
  \end{center}
 \vskip -0.80cm
 \caption{ Feynman - $x$ distributions  of the produced  $D^{0}+\bar{D^{0}}$ mesons in $pp$ collisions at 
(a) $\sqrt{s} = 13$ TeV  and 
(b) $\sqrt{s} = 200$ TeV considering different models for the intrinsic component.}
 \label{Fig:dsdxf_energy}
\end{figure}

\begin{figure}[htb]
 \begin{center}
 \subfigure[ ]{ \includegraphics[width=0.475\textwidth] 
%{razaoxf_BHPS_NEW.eps}}
{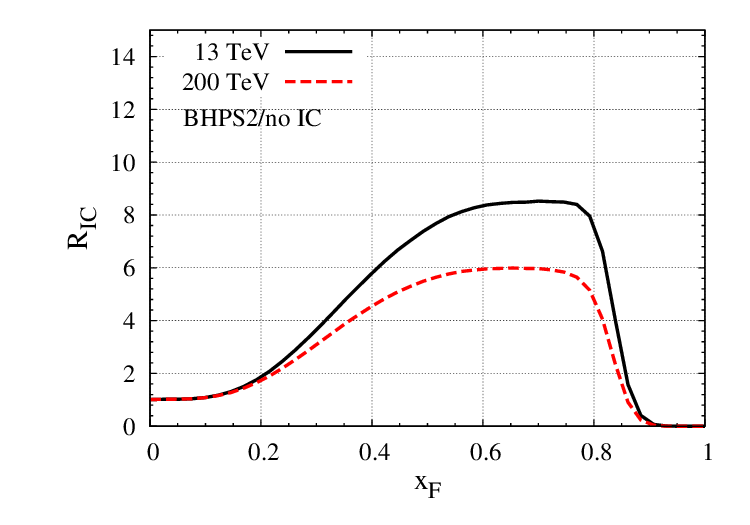}} 
\subfigure[ ]{\includegraphics[width=0.475\textwidth]
%{razaoxf_MC_NEW.eps}}
{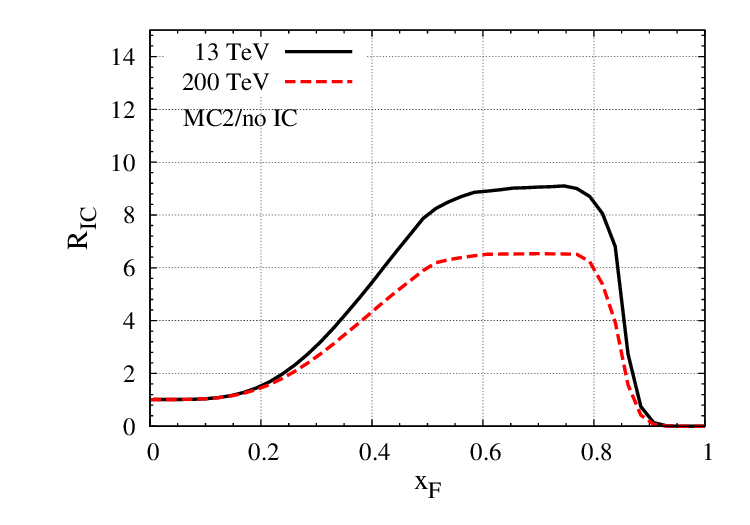}}
  \end{center}
 \vskip -0.80cm
\caption{Feynman - $x$ dependence of the ratio between the IC  predictions and the standard CTEQ 6.5 parametrization. a) BHPS2. b)MC2.} 
 \label{Fig:ratioxf}
\end{figure}

Although the  intrinsic component is predicted to manifest itself in a rapidity region beyond the current kinematical rapidity range probed by the LHC, it may also  
have  implications for other  observables. As already emphasized in Ref. \cite{npanos}, the IC component modifies the $p_T$ spectra for a fixed rapidity. In 
particular, we can access large values of $x$ in the projectile wave function by increasing the transverse momentum for a fixed $y$. However, the clear identification 
of the  intrinsic component is a hard task, since the $p_T$ spectra can also be modified  e.g. by higher order corrections. Another observable that can be affected 
by the IC component is the prompt atmospheric neutrino flux, which is strongly dependent on the features of $D$ - meson production at very forward rapidities 
(For a recent detailed discussion see Ref. \cite{anna,laha,vicantoni}). As demonstrated in Ref. \cite{vicantoni}, the main contribution to the neutrino flux comes 
from rapidities 
beyond the LHCb range, exactly where we  predict the largest impact of IC. As one of the main ingredients to calculate the prompt neutrino flux associated to the 
decay of open charm mesons is the Feynman $x$ - distribution, in what follows we will analyse in more detail the influence of the IC on this distribution for 
different energies. In Fig. \ref{Fig:termos} we show our predictions for the $x_F$ - distribution of  $D^{0}+\bar{D^{0}}$ mesons, produced in $pp$ collisions at 
$\sqrt{s} = 13$ TeV. We  present separately the gluon and charm contributions, as well as the sum of the two terms, denoted ``total'' in the figures. For comparison 
we present in Fig. \ref{Fig:termos} (a) the standard CTEQ 6.5 predictions, which are obtained disregarding a possible intrinsic charm in the initial conditions of 
the DGLAP evolution. In this case, the charm contribution is smaller than the gluonic one for all values of $x_F$, and the distribution is dominated by the 
production of $D$ mesons in gluon - gluon interactions. In contrast, when  intrinsic charm is included, the behavior of the distribution in the intermediate $x_F$ 
range ($0.2 \le x_F \le 0.8$) is strongly modified, as we can see in  Fig.  
\ref{Fig:termos} (b), where we present the BHPS2  predictions.   
 In order to analyse the energy dependence of the $x_F$ - distribution, in 
Fig. \ref{Fig:dsdxf_energy} we present our predictions for this distribution considering $pp$ collisions at  (a) $\sqrt{s} = 13$ TeV 
and (b) $\sqrt{s} = 200$ TeV.  The latter value is equivalent to the energy probed when ultra - high energy cosmic rays interact with the atmosphere. 
We have checked that 
the BHPS2 and MC2 predictions are similar for all energies considered.  
 In order to determine the magnitude of the impact 
of the IC and the kinematical range influenced by its presence, we present in Fig. \ref{Fig:ratioxf} (a) and (b) our predictions for the ratio between the $x_F$ 
distributions predicted by the BHPS and MC models and the standard CTEQ 6.5 one.  As expected from Fig. \ref{Fig:dsdxf_energy}, the BHPS and MC models predict an 
enhancement at intermediate $x_F$ and a suppression at very large $x_F$. Moreover, the magnitude of the enhacement is similar for both models, being a factor 
6 -- 9 in the energy ranges considered. The main  aspect that should be emphasized here is that {\it the enhacement occurs exactly in the $x_F$ range where the 
contribution of  $D$ - mesons to the prompt neutrino flux \cite{anna,laha,vicantoni} is dominant}. Consequently, we expect that the presence of the IC component 
will modify the predictions for the prompt atmospheric neutrino flux. This expectation will be analysed in detail in a future publication.

\section{Conclusions}
\label{conc}
A complete knowledge of the partonic structure of  hadrons is fundamental to make predictions for the Standard Model and  beyond Standard Model processes at 
hadron colliders. In particular, the heavy quark contribution to the proton structure has a direct impact on several observables analysed at the LHC. Direct 
measurements of heavy flavors in DIS and hadronic colliders are consistent with a perturbative  origin. However, these experiments are in general not sensitive 
to heavy quarks at large $x$. Therefore, it is  fundamental to propose and study other observables which may be used to determine the presence (or absence)  of an 
intrinsic heavy quark component in the hadron wave function. In recent years, a series of studies have discussed in detail how to  
probe this intrinsic component, with particular emphasis in processes that are strongly sensitive to the charm in the initial state. One of this processes is   
$D$ - meson production at forward rapidities, which is also influenced by the specific features of QCD dynamics at high energies. In this paper we have extended 
the approach proposed in Ref. \cite{npanos} to the production of $D$ - mesons from charm quarks present in the initial state and we  have calculated the rapidity 
and $x_F$ - distributions of $D$ - meson produced in $pp$ collisions at the LHC and in 
interactions at higher energies. In particular, we have included the contribution associated 
to gluon - gluon interactions, which are also affected by the intrinsic charm component. Considering different models of the intrinsic charm component, we have 
demonstrated that the rapidity range influenced by IC is beyond that reached by the LHCb Collaboration. However IC is important for the calculation of the prompt 
neutrino flux. Our results indicated that the $x_F$ - distribution is enhanced by the intrinsic component in the kinematical range that dominates the $D$ - meson 
contribution to  the prompt neutrino flux. Consequently, the inclusion of the IC contribution in the corresponding calculations can be important to estimate the 
prompt atmospheric neutrino flux probed by the ICECUBE.

%-------------------------------------------------------------------------------
\section*{Acknowledgments}
VPG thanks Anna Stasto, Roman Pasechnik, Antoni Szczurek and Rafal Maciula for useful 
discussions about the $D$ - meson production at high energies and its implications on 
the prompt neutrino flux. FSN is grateful to Juan Rojo and J.P. Lansberg for useful discussions.
This work was partially financed by the Brazilian funding agencies CAPES, CNPq,  FAPESP and FAPERGS.

\end{document}